# Pressure, stress, and strain distribution in the double-stage diamond anvil cell

Sergey S. Lobanov[1,2,*], Vitali B. Prakapenka[3], Clemens Prescher[3], Zuzana Konôpkova[4], Hanns-Peter Liermann[4], Katherine Crispin[1], Chi Zhang[5], Alexander F. Goncharov[1,6,7]

[1]Geophysical Laboratory, Carnegie Institution of Washington, Washington, DC 20015, USA

[2]V.S. Sobolev Institute of Geology and Mineralogy SB RAS, Novosibirsk 630090, Russia

[3]Center for Advanced Radiation Sources, University of Chicago, Chicago, IL 60632, USA

[4]Photon Sciences DESY, D-22607 Hamburg, Germany

[5]Key Laboratory of Earth and Planetary Physics, Institute of Geology and Geophysics CAS, Beijing 100029, China

[6]Key Laboratory of Materials Physics, Institute of Solid State Physics CAS, Hefei 230031, China

[7]University of Science and Technology of China, Hefei 230026, China

*slobanov@carnegiescience.edu

## Abstract

Double stage diamond anvil cells (DAC) of two designs have been assembled and tested. We used a standard symmetric DAC as a primary stage and CVD microanvils machined by a focused ion beam – as a second. We evaluated pressure, stress, and strain distributions in Au and Fe-Au samples as well as in secondary anvils using synchrotron x-ray diffraction with a micro-focused beam. A maximum pressure of 240 GPa was reached independent of the first stage anvil culet size. We found that the stress field generated by the second stage anvils is typical of conventional DAC experiments. The maximum pressures reached are limited by strains developing in the secondary anvil and by cupping of the first stage diamond anvil in the presented experimental designs. Also, our experiments show that pressures of several megabars may be reached without sacrificing the first stage diamond anvils.

**Introduction**

The invention of diamond anvil cell (DAC) in 1959 (Ref.[1,2]) made it possible to explore extreme energy-density regimes[3-5]. In particular, DACs found many applications in geophysics and planetary sciences as it allow simulating pressure and temperature conditions of planetary interiors in the lab[6]. Reaching extreme pressures, however, is very challenging, and often pressure itself cannot be accurately determined. The highest possible pressure that can be reached using DACs has been a matter of intense debate[7,8]. In order to reach higher pressures, the design of DACs went through a number of development iterations such as the introduction of beveled diamond anvils[9] and other refinements of the anvil shape[10-12] leading eventually to a multi-beveled anvil geometry[13]. In addition, anvil stability was enhanced by the use of synthetic defect-free single crystal diamonds[14]. Most recently, the synthesis of mechanically isotropic nano-diamonds that hamper premature diamond failure due to the absence of the weak cleavage[15] has improved the anvil stability and extended the accessible pressure range. Further improvements came from the recognition that gasket's high yield strength and high ductility is of primary importance[10]. These DAC developments have been performed by a combination of experimental trial and error method as well as finite element calculations[10,11] and allowed reaching pressures up to ~400 GPa[13].

The recent work by Dubrovinsky *et al.* [16] reported pressures in excess of 600 GPa indicating that DAC pressure limits can be substantially increased when using nano-crystalline diamonds as second stage anvils. Although double stage anvil technique sets the stage for further high pressures studies, the experimental success rate was reported to be low due to the alignment difficulties and sample gliding[16]. Also, it remains unclear what causes the substantial pressure increase as well how such high pressures are sustained. Is it the exceptional mechanical properties of nano-diamonds, which are entirely responsible for such an improvement? Is it the introduction of a second stage anvil that is important for reaching such high static pressures? Careful characterization of pressure gradients and stress-strain relations in the secondary diamond anvils and in the gasket may allow answering these questions and provide an important ingredient for reproducibly reaching pressures in excess of 400 GPa, beyond the limit of conventional DAC technique.

Current progress in material machining using focused ion beam (FIB) technology enables the fabrication of secondary anvils of a variety of shapes and dimensions, which may assist the reproducibility of the double stage DAC technique. Sakai *et al.* [17] have shown that pressures over 300 GPa may be generated when using a pair of microanvils machined from a single diamond block via FIB. Microanvils were connected with a silicon rod to preserve the alignment. In our contribution we further explore how different microanvil geometries produced under a well-controlled FIB milling affect pressure generation in a double-stage DAC. Two principally different second stage diamond anvil (SSDA) geometries were tested, referred here as type-1 and type-2; both reaching pressures up to 240 GPa. Also, we address the unclear strain-stress distributions in the double-stage DAC to further optimize this device and determine its limitations.

## Methods

**General diamond anvil cell assemblage.** In the type-1 assemblage (Fig. 1A), single crystal diamonds with flat culets of 300 µm were used as first stage diamond anvils (FSDA). A shallow (~2 µm) pit (55 µm) was milled with a focused ion beam in the center of each anvil culet to ease the alignment and to add stability to SSDA. The gasket was made from a 266 µm thick Re foil indented to the thickness of 60-65 µm. The cullet area of the gasket was drilled out to be filled by cBN mixed with epoxy and compressed to 30-35 GPa between the FSDA. Then, a 55 µm hole was drilled in the center of the cBN gasket to create a chamber for SSDA and a sample. Microanvils (18-20 µm thick, 50 µm in diameter, 15 µm culets) were positioned in the pits, aligned with each other under a microscope, and glued to the FSDA with epoxy. Gold (>99.9), and iron (>99.9) powders (<10 µm) were mixed and put directly in the sample chamber with no pressure medium. In the type-2 assemblage (Fig. 1B), beveled diamonds with 300/100 µm culets without pits were used as FSDA. Re gaskets (25 µm thick) were laser-drilled in the center to create a 35 µm chamber for SSDA and a sample. SSDA (4-5 µm thick, 30 µm in diameter, 25µm culets) were positioned on the top of FSDA without gluing. Because our microanvils are transparent, it was possible to align the SSDA under the microscope in transmitted light. Sample was a 10 µm thick gold foil. Nitrogen gas was loaded at ~ 0.2 GP to serve as a pressure medium.

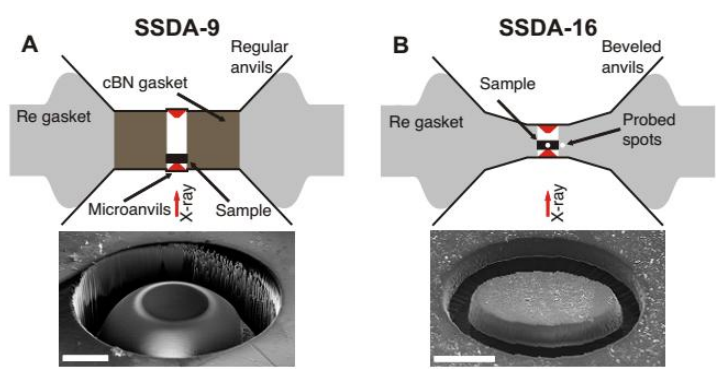

**Figure 1.** Type-1 (**A**) and type-2 (**B**) DAC assemblage (top) and SEM micrographs of the SSDA (bottom) in the CVD substrate before placing on the first stage anvils. White bars in SEM images correspond to 10 µm.

**Microanvil preparation.** 20 and 5 µm thick microcrystalline CVD diamond wafers (Diamond Materials GmbH) with a grain size of ~ 10 nm to 50 µm depending on the distance from the seed[18,19] were used to cut SSDA of desired and reproducible geometries with a focused ion beam (FIB/SEM Zeiss Auriga 40) available in the Geophysical Laboratory Carnegie Institution of Washington. Orthogonal geometry water-assisted milling was performed with the Ga ion dose of ~7 nC/µm$^3$ at the beam current of 4-16 nA. Nanofabricating was operated in the NPVE FIBICS software. After the milling microanvils were extracted from the substrate and cleaned in an ultrasonic bath in isopropanol.

**Synchrotron X-ray diffraction** was used to probe samples upon compression. Experiments were performed at the Extreme Conditions Beamline P02.2, DESY (Germany) and at the GeoSoilEnviroCARS, APS (USA). X-ray beams with energies of 42.857 keV (DESY) and 37.077 keV (APS) focused to a 2 and 3 µm spot size (FWHM), respectively. Pressure was gradually increased using membranes and x-ray diffraction (XRD) was collected at each membrane force increment. Pressure gradients were revealed by x-ray mapping at the highest membrane load. Pressures were calculated with the equations of state (EOS) for Au[20], Fe[21], Re[22], and diamond[23].

**Results and Discussion**

Below we describe two experiments SSDA-9 (DESY) and SSDA-16 (GSECARS) representative of the type-1 and type-2 assemblages, respectively. Relatively large number of experiments were successful (~90 %, P > 150 GPa) independent of the initial gasket thickness, SSDA geometry and positioning, or sample composition. Second stage anvil alignment, however, was crucial to reach pressures above 200 GPa.

Figure 2 shows XRD patterns in the SSDA-9 experiment at the highest membrane load as a function of distance from the sample center (found by x-ray transmission profiles, also corresponds to the culet center). Pattern **A** was collected 15 µm away from the sample center and both the EOS of Au and Fe indicate P = 71 GPa based on the positions of the diffraction lines. Gold and iron form the most intense peaks and a texture typical for powder samples (vertical lines with uniform intensity distribution along the azimuth) in the diffraction images (Fig. 3). Diffraction lines in **A** and **B** show clear waviness which is an indication of stress (Fig. 3) and is not typical in experiments with axial XRD geometry as only the least stressed crystallites satisfy the Bragg's law[24]. The sinusoidal diffraction lines in **A-B** patterns indicate that the stress field is not uniaxial and pressure gradients exist along the compression axis near the microanvil edge. On the other hand, patterns **C-F** lack waviness implying that in the sample center stress field may be considered uniaxial. Bragg reflections from polycrystalline microanvils form a spotty distribution in the XRD images and are easy to identify.

Significant changes take place in the **A** to **D** (15 to 6 µm away from the sample center) XRD patterns while approaching the SSDA even though pressure remains nominally constant (71 GPa). The intensity of the gold and iron Bragg peaks decreases, while the diamond 111, 220, and 311 reflections start splitting indicating that SSDA are highly strained (Fig. 3). 6 µm away from the sample center (pattern **D**) a shoulder appears at $2\theta = 8.1°$ which cannot be attributed to the Fe 100 peak ($2\theta = 8.15°$) and further intensifies in patterns **E** (3 µm) and **F** (0 µm) (Fig. 2). This new reflection shows a powder-like uniform intensity distribution along the diffraction azimuth (Fig. 3) suggesting it is the Au 111 reflection, but at P = 216 GPa. The Au 220 peak shows a similar discontinuous behavior in the **A-F** patterns and yields consistent unit cell volumes. Positions of iron diffraction lines also experience a discontinuous change at 6 µm away from the sample center. A new strong peak appears at $2\theta = 9.9°$ in the **D-F** patterns (Fig. 2). Based on the homogeneous intensity distribution along the diffraction azimuth (Fig. 3) we assign this new peak to the Fe 101 line, yielding $P_{Fe}$ = 216 GPa. Unfortunately, the Fe 002 and Au 200 peaks overlap and we were not able to reliably resolve them at high pressure. The relatively weak Fe 100 peak is seen in XRD images, but is dominated by SSDA reflections in the integrated XRD patterns. Pressures yielded by Au and Fe EOS are consistent within the uncertainty of ±3 GPa (2 sigma) based on the refinements of the Au and Fe unit cell. Interestingly, noticeable signal of gold and iron at P~70 GPa is observed 3 and 0 µm away from the sample center (patterns **E** and **F)** indicating that the materials are also probed aside from the microanvils due to the 'tails' of the x-ray beam. A similar behavior was recently observed by Sakai *et al.* [17] implying that highly-focused x-ray beams are necessary for double-stage DAC studies.

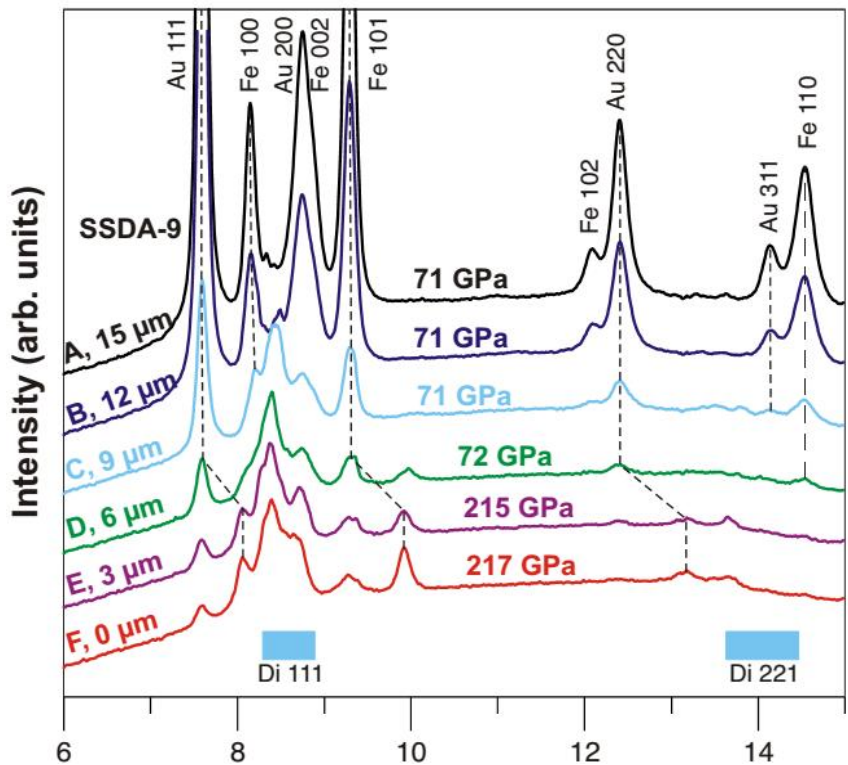

**Figure 2.** XRD patterns of SSDA-9 at ~60 bar membrane pressure collected with a 3 μm step. Pattern **A** is 15 μm away from the sample center and pattern **F** corresponds to the sample center. As determined from the XRD image analysis (Fig. 3), Au 111 and 220 as well as Fe 101 Bragg peaks show a discontinuous shift from pattern **D** to **E** (marked with black dashed lines). Au 200 and Fe 002 peaks overlap and are not clearly resolved in **E** and **F** patterns. Blue bars correspond to the 2θ range where diamond 111 and 220 reflections are expected in the 71-217 GPa range[23]. X-ray wavelength is 0.2893 Å.

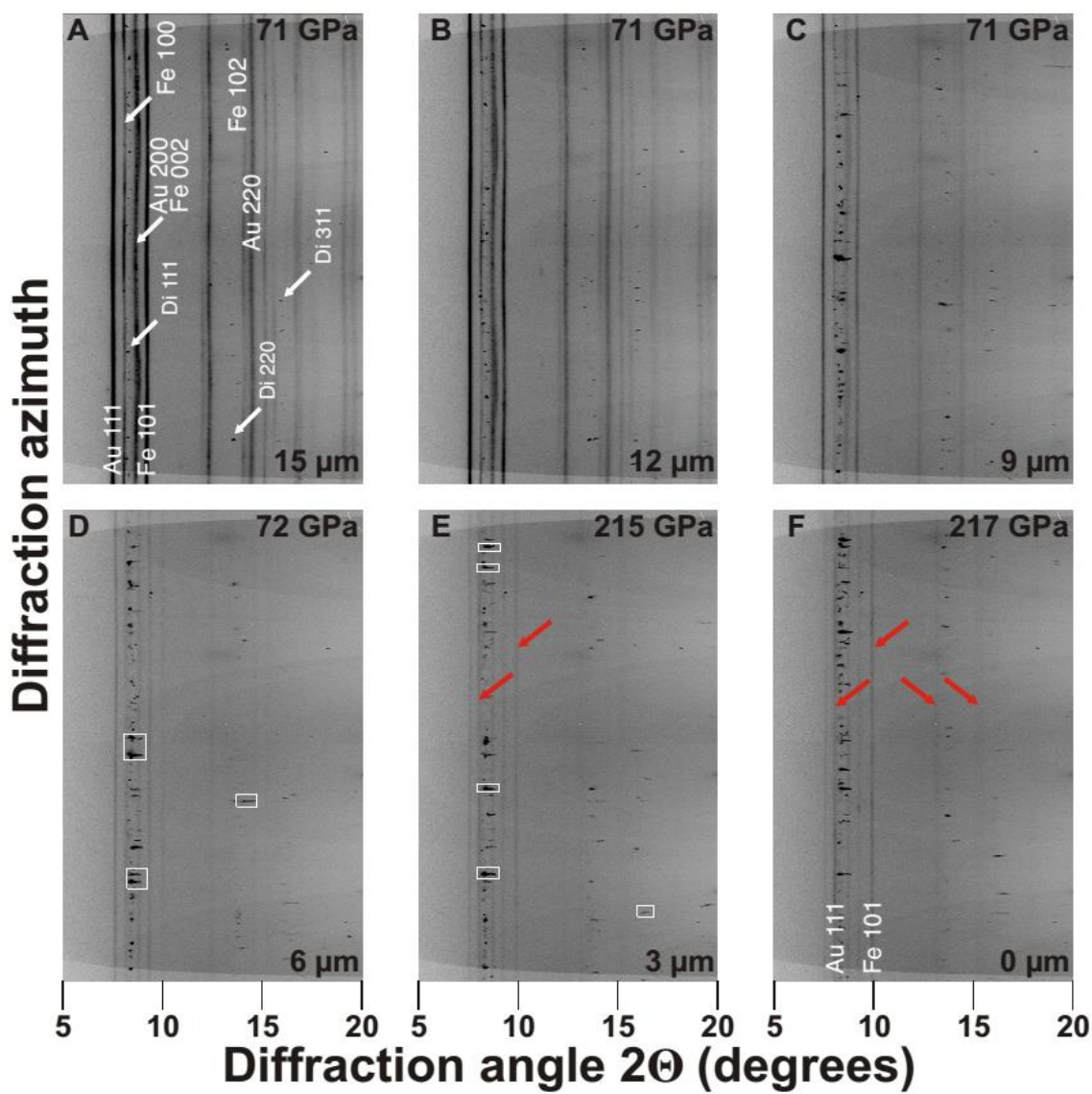

**Figure 3.** Caked SSDA-9 XRD as a function of distance from the sample center. Red arrows depict new Au and Fe Bragg peaks appearing in XRD collected near the sample (culet) center. White arrows point to some of the diamond, gold, and iron Bragg reflections. White rectangles mark severely split diamond reflections. The waviness of diffraction lines in patterns A-B indicates pressure gradients along the probing direction. Note that the patterns notation here corresponds to that in Fig. 2. X-ray wavelength is 0.2893 Å.

Pressure gradients were revealed by x-ray mapping the sample with a 3 μm step. Both gold and iron were used to reconstruct the pressure distribution. The resulting maps show extremely steep pressure gradients right on the culet edge of the SSDA reaching approximately 50 GPa/μm (Fig. 4).

The circular dome-like feature is indicative of a good alignment of the SSDA. In a number of other runs with the same SSDA assemblage we observed elliptical domes in the pressure maps suggesting that microanvils probably were not perfectly aligned initially or got misaligned upon compression. Intriguingly, pressure values in the very center of the dome seem slightly lower than that on the microanvil edge. One interpretation could be that the microanvils are cupping, a well know phenomenon in conventional DAC experiments[9,25-28].

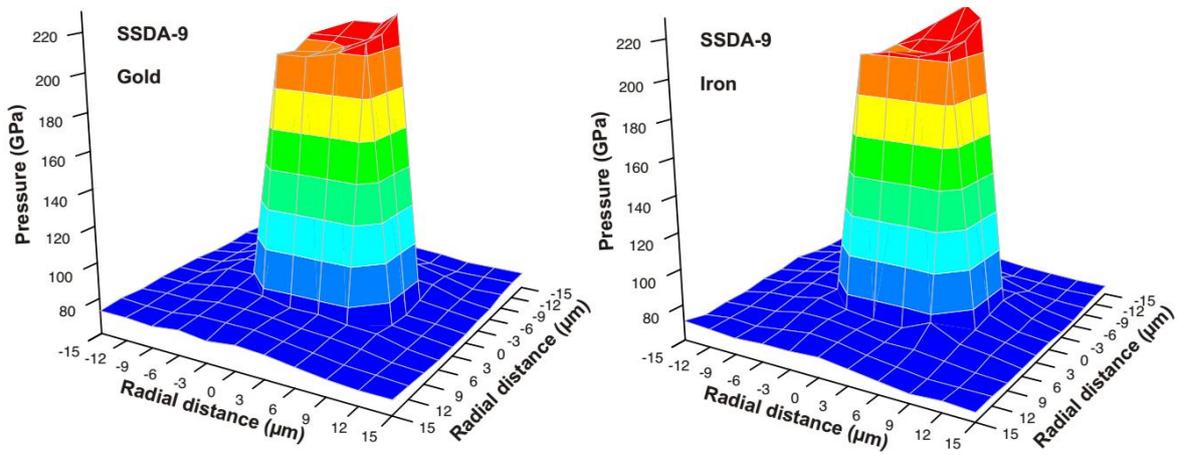

**Figure 4.** Pressure distribution in the SSDA-9 at the membrane load of 60 bar.

Split diamond reflections contain information on the lattice strains accommodated by a single diamond grain in the SSDA (Fig. 3) through $\epsilon(hkl) = \frac{d_m(hkl) - d_P(hkl)}{d_P(hkl)}$ (Ref.[29]) where $d_m(hkl)$ is the lowest *d*-spacing value of a grain, and $d_P(hkl)$ is the highest *d*-spacing value of the same grain. We have analyzed 4 diffraction images with the most pronounced splitting and derive $<\epsilon_{111}>$ = -0.059, $<\epsilon_{220}>$ = -0.057, $<\epsilon_{311}>$ = -0.047, where $<\epsilon_{hkl}>$ denote averaged strains perpendicular to a given *hkl* plane. These numbers are consistent with the cleavage energies increasing from *111* to *110* to *311* (Ref.[30]) as well as with the directional dependent Hugoniot elastic limits of diamond[31]. Pressure differences for individual split reflections reaches 200 GPa, which is higher than the yield strength of single crystal diamond (130-140 GPa)[32], suggesting that SSDA may be plastically cupped. The use of SSDA that are isotropic on a nanometer scale[15] may allow extending the pressure limit of the double stage DAC technique because of the absence of the weak cleavage direction.

X-ray transmission profiles as well as the diminishing intensities of all Brag peaks near the sample center suggest a strong material flow away from the SSDA center. One possible reason is the lack of gasket support for the sample on the microanvil edge where the maximum pressure gradients occur. In order to decrease pressure gradients near the microanvil edge and reach higher pressures we have tested a 300/100 μm beveled diamond anvils that themselves may be used to reach P ~ 150 GPa (Ref.[33]). We decreased the SSDA anvil height as it may also assist in reducing the pressure gradients in a similar manner to anvils with small bevel angles (< 8.5°) producing less steep pressure distributions[12,34]. Additionally, we increased the SSDA culet size to 25 μm for a better sample support as will be now shown in the SSDA-16 experiment.

In the SSDA-16 run we gradually increased membrane load to 60 bar while probing pressure in the sample center (gold and diamond) and 30 μm aside (rhenium) (Fig. 1B). Variations of pressure obtained for gold, diamond, and rhenium are shown in Figure 5 as a function of membrane load. Not surprisingly, gold in the sample center yields the maximum pressures, while diamond shows systematically lower pressures close to that observed 30 μm aside. Figure 6 shows the differences in pressure between the probed spots and underscores the role of microanvils in generating high pressure. Interestingly, at membrane load above 20 bar the pressure differences remain almost constant. A disintegration of the SSDA at ~ 20 bar or, alternatively, an onset of the anvil cupping may be the cause. The number of spotty diamond reflections observed in the XRD images did not change over the entire compression cycle as one would expect in the case of SSDA breakdown. Instead, the SSDA sustained large stresses as evidenced by the maximum strains observed: $\epsilon_{111}$ = -0.023, $\epsilon_{220}$ = -0.014. Also, at the membrane load of 20 bar gold yields P ~ 140 GPa in the sample center, which is close to the diamond yield strength (130-140 GPa)[32] favoring the cupping scenario.

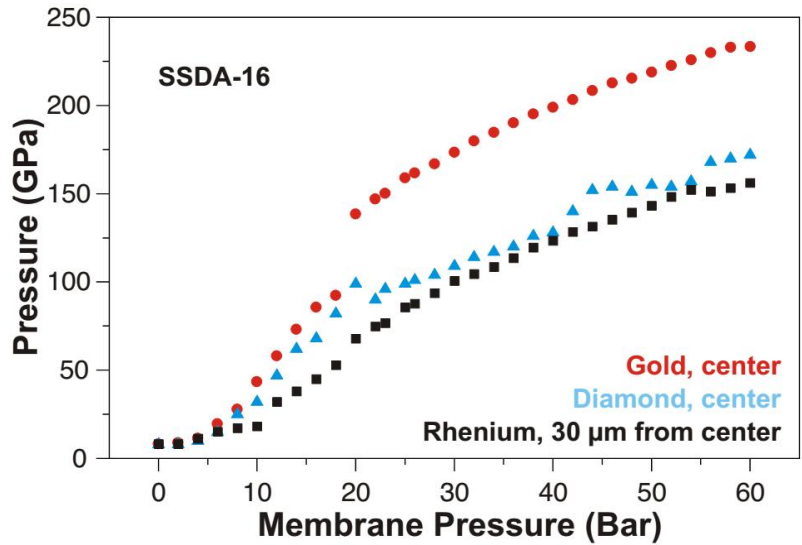

**Figure 5.** Pressure vs membrane load. $P_{Au}$ and $P_{Di}$ were measured in the same central spot, while $P_{Re}$ was measured 30 μm away from the sample center.

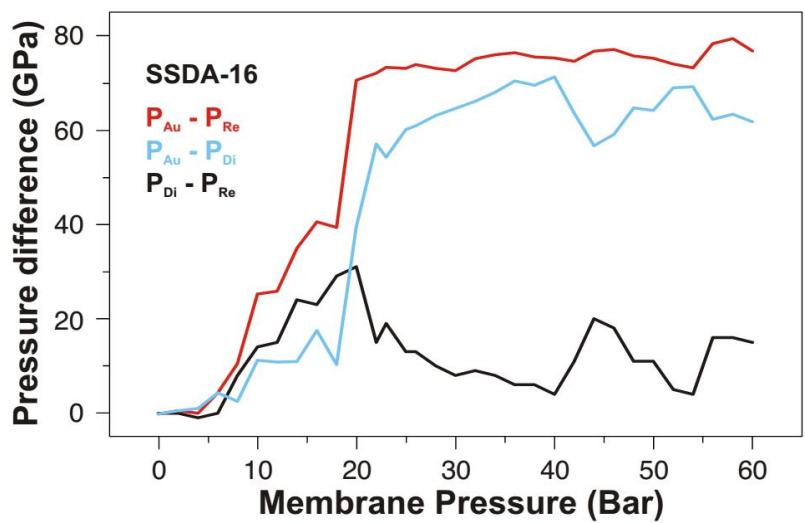

**Figure 6.** Difference in pressures obtained in the sample center (gold and diamond) and 30 μm aside (rhenium) vs membrane load.

In order to reveal the pressure distribution in detail we performed a one-dimensional XRD scan across the sample chamber at the maximum membrane load of 60 bar. Figure 7 shows four

XRD patterns collected towards the sample center starting 13 μm aside with a step of 2 μm. The maximum pressure gradients are concentrated at the microanvil edge and are approximately 22 GPa/μm (Fig. 8A), which is ~ 2 times smaller than that in the SSDA-9 experiment. Type-2 assemblage allows preserving more sample material in between the microanvils as evidenced by four distinct gold peaks in patterns C and D (9 and 7 μm away from the sample center, respectively). Further approaching the sample center did not change diffraction patterns.

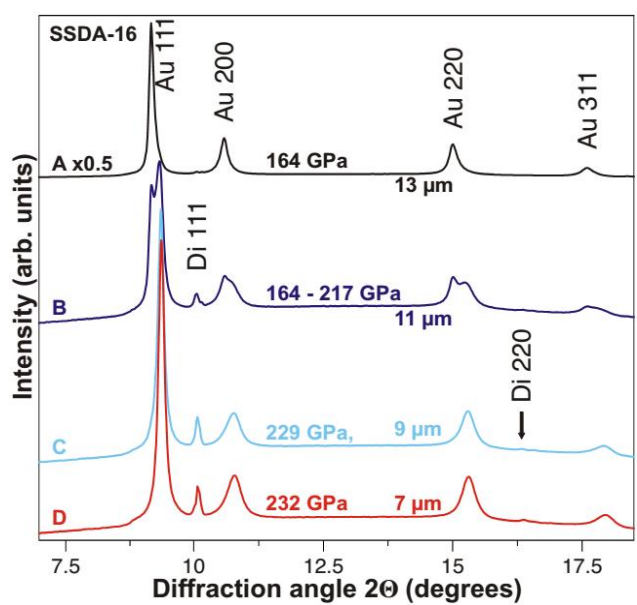

**Figure 7.** XRD patterns of SSDA-16 at ~60 bar membrane load collected with a 2 μm step starting 13 μm away from the sample center. Patterns **A**, **B**, **C**, and **D** are 13, 11, 9, and 7 μm away from the sample center, respectively. X-ray wavelength is 0.3344 Å.

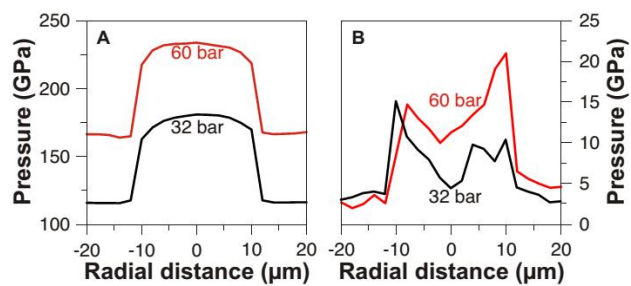

**Figure 8.** (**A**) Pressure gradients across the sample chamber at the membrane load of 32 and 60 bar. (**B**) Difference in pressures between values estimated based on Au 111 and 200 Bragg peaks.

Inasmuch as under non-hydrostatic conditions the measured unit cell parameter varies as a function of *hkl*, the variation of $P_{111} - P_{200}$ across the sample chamber may serve as a qualitative measure of the stress spatial variation, where $P_{111}$ and $P_{200}$ are pressures determined solely from the positions of gold 111 and 200 Bragg peaks. The maximum $P_{111} - P_{200}$ values correspond spatially to the microanvil edges (Fig. 8B), where the stress field is likely complex and cannot be accurately defined. Nevertheless, reducing the magnitude of nonhydrostaticity at the microanvil edge by optimizing FSDA or SSDA geometry may assist reaching higher pressures.

In the sample center the stress field may be assumed to be uniaxial[35] and, by following the previously developed formalism[13,35,36], one may calculate the uniaxial stress component, *t*. For the cubic system, the collected XRD data may be represented in the linear form:

$$a_m(hkl) = M_0 + M_1[3\Gamma(hkl)(1 - 3\sin^2\theta_{hkl})],$$

$$\Gamma(hkl) = (h^2k^2 + k^2l^2 + h^2l^2)/(h^2 + k^2 + l^2)^2,$$

where *hkl* are Miller indices of a given plane and $\theta_{hkl}$ its diffraction angle ($2\theta_{hkl}/2$). Plotting $a_m(hkl)$ versus $3\Gamma(hkl)(1-3\sin^2\theta_{hkl})$ for the observed diffracting planes *hkl* allows solving for $M_0$ and $M_1$. We now obtain the uniaxial stress component using:

$$t \cong -3M_1/\alpha S M_0,$$

$$S = S_{11} - S_{12} - S_{44}/2,$$

here, the parameter α, which varies from 0 to 1, is a measure of stress and strain continuity across grains in a polycrystalline sample, $S_{ij}$ are the gold single-crystal elastic compliances[37] calculated using the Birch's finite strain theory[38]. Pressure was calculated from the position of gold *111* peak, as gold *hhh* reflections are the least sensitive to nonhydrostatic conditions[39]. Isothermal elastic constants of gold and their pressure derivatives were taken from Hiki and Granato [40] and α was assumed 1 to allow comparison with previously reported values[41,42]. The obtained $t$ values moderately depend on the assumed $C_{ij}$ and their pressure dependencies. For example, using elastic constants of gold and their pressure dependencies reported in Golding *et al.* [43] yields values up to 0.5 GPa higher than that obtained when using the constants from Hiki and Granato [40]. Uniaxial stress varies as a function of pressure and may be compared with $t$ values reported in earlier studies (Fig. 9). Interestingly, uniaxial stresses derived in this study (axial XRD, without pressure medium) are systematically larger than the values reported in a study by Dorfman *et al.* [42], where gold samples were loaded with Ne pressure medium. On the other hand, the pressure dependence of $t = 0.06 + 0.015P$ determined in a study with radial XRD geometry and without pressure medium[41] predicts larger uniaxial stress values (Fig. 9), which is consistent as radial XRD allows probing the most stressed crystallites[24].

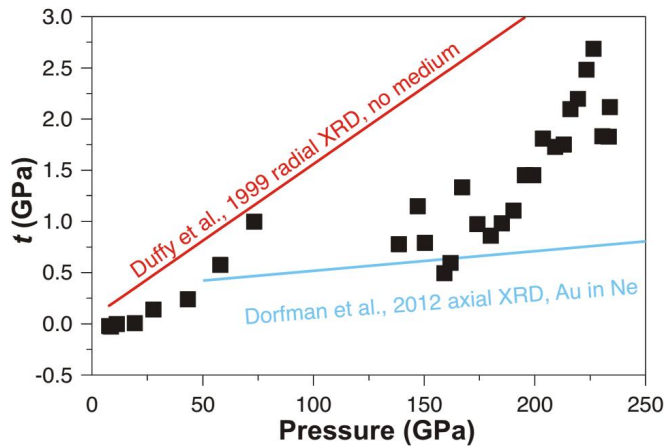

**Figure 9.** Uniaxial stress component of gold (black squares, present data) as a function of pressure (measured via axial XRD). Red line, $t = 0.06 + 0.015P$, represents gold uniaxial stress values as determined by radial XRD (Ref.[41]). Blue line is the guide to the eye through the uniaxial stress values of gold in Ne pressure medium as determined by axial XRD (Ref.[42]).

To summarize, the experimental assemblages of double stage DAC technique used in this study allow routinely reaching pressures in excess of 200 GPa. FIB proved useful in manufacturing

SSDA of desired geometry which helped to increase the success rate of the double-stage technique to over 90% in the case of this study. SSDA crafted from CVD diamonds underwent deformation, which likely limited the maximum pressures reached in this work. Importantly, tailoring microanvils from nanocrystalline diamonds may help to routinely reach pressures of several megabars. Additionally, higher pressure may be achieved through modifying the experimental geometry based on the analysis of nonhydrostaticity distribution across the high-pressure area as shown in this work and modifying the experimental geometry to optimize the stress concentrations. Microanvils used in this study are transparent in the visible range SSDA and are perfectly suitable for vibrational and optical spectroscopy studies as well as for laser-heating experiments at $P > 200$ GPa.

## Acknowledgement

We thank S. Sinogeikin (HPCAT) for measuring the pressure map on the initial stage of this work. This work was supported in part by the Deep Carbon Observatory and NSF EAR 1015239, EAR 1128867. A.F.G. acknowledges the support of NSFC, the grant number is 21473211. S.L. would like to thank the support of Ministry of Education and Science of Russian Federation (No 14.B25.31.0032). Parts of this research were carried out at the light source PETRA III at DESY, a member of the Helmholtz Association (HGF). Portions of this work were performed at GeoSoilEnviroCARS (Sector 13), Advanced Photon Source (APS), Argonne National Laboratory. GeoSoilEnviroCARS is supported by the National Science Foundation - Earth Sciences (EAR-1128799) and Department of Energy- GeoSciences (DE-FG02-94ER14466). This research used resources of the Advanced Photon Source, a U.S. Department of Energy (DOE) Office of Science User Facility operated for the DOE Office of Science by Argonne National Laboratory under Contract No. DE-AC02-06CH11357.